\documentclass[epj]{svjour}
\usepackage{graphicx}
\usepackage{amsmath,amssymb}
\usepackage{color}

\begin{document}
\sloppy

\title{Monte Carlo simulations of a disordered superconductor-metal quantum phase transition}
\titlerunning{Disordered superconductor-metal transition}

\author{Ahmed K. Ibrahim \and Thomas Vojta}

\institute{Department of Physics, Missouri University of Science and Technology, Rolla, Missouri 65409, USA}
	
\date{Received:  / Revised version: }

\abstract{
We investigate the quantum phase transitions of a disordered nanowire from superconducting to metallic behavior by employing extensive Monte Carlo simulations. To this end, we map the quantum action onto a (1+1)-dimensional classical XY model with long-range interactions in imaginary time. We then analyze the finite-size scaling behavior of the order parameter susceptibility, the correlation time, the superfluid density, and the compressibility. We find strong numerical evidence
for the critical behavior to be of infinite-randomness type and to belong to the random transverse-field Ising universality class, as predicted by a recent strong disorder renormalization group calculation.
}


\maketitle


\section{Introduction}
\label{sec:Intro}

Investigating the electrical transport in low-dimensional fluctuating superconductors has attracted great interest during the last decades. Early experiments \cite{WebbWarburton68} demonstrated nonzero resistivity below the bulk critical temperature. This was explained using the notion of thermally activated phase slips \cite{Little67,LangerAmbegaokar67,McCumberHalperin70}. Later, quantum phase slips were considered as well \cite{Giordano88,ZGOZ97}. Recent experiments \cite{LMBBT01,BVBK04,RogachevBollingerBezryadin05,ACMHT06,RWPBGB06} studied the electrical transport characteristics of one-dimensional ultrathin metallic nanowires.  Measurements of the resistance demonstrated that thicker wires undergo a phase transition from a metallic state to a superconducting state upon decreasing the temperature. However, thinner wires do not show superconductivity even at the lowest temperatures $T$.

The behavior of these experiments can be understood using the concept of a superconductor-metal quantum phase transition (as a function of wire thickness) in the pair-breaking
universality class, as proposed in recent theoretical work \cite{SachdevWernerTroyer04,DRSS08,DelMaestroRosenowSachdev09}. The pair breaking is likely caused by magnetic moments on the wire surface. The nanowires in question are narrower than the (bulk) superconducting coherence length but still contain
a large number of transverse channels for unpaired electrons. The resulting theory is therefore built on a model of one-dimensional superconducting fluctuations whose dynamics is overdamped because of the coupling to three-dimensional unpaired electrons
\cite{Herbut2000,LopatinShahVinokur05,SpivakTyuzinHruska01,FeigelmanLarkinSkvortsov01,Galitski08}.

Quenched disorder plays a significant role in these nanowires due to the random positions of the pair-breaking magnetic moments. The thermodynamics of the resulting disordered superconductor-metal quantum phase transition has been studied analytically via a strong-disorder renormalization group analysis \cite{HoyosKotabageVojta07,VojtaKotabageHoyos09} and numerically
\cite{DRMS08} via a solution of the saddle-point equations of the corresponding Landau-Ginzburg-Wilson (LGW) order parameter field theory. Both methods invoke the large-$N$ limit in which the number of order parameter components is generalized from 2 (representing the real and imaginary parts of the Cooper pair density) to $N\gg 1$. These methods predict that the quantum phase transition is governed by an unconventional nonperturbative infinite-randomness critical point in the same universality class as the random transverse-field Ising model
\cite{Fisher92,Fisher95}.
Its dynamical scaling is of activated rather than power-law type, i.e., the correlation time $\xi_\tau$ is related to the correlation length $\xi$ as $\text{ln} \xi_\tau \sim \xi^\psi$, where $\psi=0.5$ is the tunneling exponent.
Observables also show nonconventional scaling behavior. For example, the order parameter susceptibility diverges not just at criticality
but in an entire parameter region, the quantum Griffiths phase, around the transition. The superfluid density on the superfluid side of the transition also behaves anomalously. It remains zero in part of the long-range ordered quantum Griffiths phase. As these results have been obtained within the large-$N$ approximation, it is important to verify that they remain valid for the physical case of a two-component order parameter.

In this paper, we therefore investigate the effects of disorder on the quantum phase transition between superconductor and metal in thin nanowires by employing a Monte Carlo method. This allows us to test the predictions directly for $N=2$ order parameter components. Our paper is organized as follows: In Sec.\ \ref{sec:Model}, we define the overdamped Cooper pair model and describe the mapping onto a classical XY Hamiltonian. In Sec.\ \ref{sec:Theory}, we briefly summarize the renormalization group predictions. Section \ref{sec:MC} introduces the Monte Carlo simulations. In Sec.~\ref{view}, we discuss our results and compare them to the predictions of the strong-disorder renormalization group. We conclude in Sec.\ \ref{sec:Conclusions}.

\section{The model}
\label{sec:Model}
The starting point of our work is a quantum LGW order parameter field theory (i.e., free energy functional) for an $N$-component vector order-parameter $\varphi =(\varphi_1,...,\varphi_N)$ in $d$ space dimensions. (We will later set $d=1$ and $N=2$ as appropriate for the superconductor-metal transition in nanowires where $\varphi$ represents the Cooper pair density.) The LGW action can be derived from a Hamiltonian of disordered electrons by employing standard techniques
\cite{Hertz76,Millis93,KirkpatrickBelitz96,BelitzKirkpatrickVojta05}. In the absence of quenched disorder, the action reads
\cite{HoyosKotabageVojta07,DRHV10}
\begin{equation}
S=\int dydx \, \varphi(x) \Gamma(x,y) \varphi(y)+ \dfrac{u}{2N} \int dx \, \varphi^4(x),\label{qaction}
\end{equation}
where $x\equiv (\textbf{x},\tau)$ is a vector that includes position $\textbf{x}$ and imaginary time $\tau$, $\int dx \equiv \int d^d\textbf{x} \int d\tau$, and $u$ is the standard quartic coefficient. The Fourier transform of the bare inverse propagator (two-point vertex) $\Gamma (x,y)$ reads
\begin{equation}
\Gamma (\textbf{q},\omega_n)=r+\xi_0^2 \textbf{q}^2+\gamma_0 |\omega_n|^{2/z_0} .
\end{equation}
Here, $r$ denotes the distance from criticality, $\xi_0$ is a microscopic length,  $\omega_n$ represents the Matsubara frequency, and $\gamma_0$ is the damping coefficient. For the Ohmic order parameter dynamics caused by the coupling to the conduction electrons, the value of the bare
dynamic exponent is $z_0=2$. Quenched disorder can be introduced into the action (\ref{qaction}) by making the distance from criticality
(and/or the other coefficients) a random function of real-space position, $r \to r+\delta r(\mathbf{x})$.

In preparation for the Monte Carlo simulations, we now set $d=1$, $N=2$ and map the quantum action  onto a ($1+1$)-dimensional classical XY model. This can be accomplished by discretizing space and imaginary time in the action (\ref{qaction}) and interpreting imaginary time as another space dimension. As a result, the classical XY Hamiltonian reads:
\begin{eqnarray}
H&=&-\sum_{i,\tau} (J_i^{(s)} S_{i,\tau} S_{i+1,\tau}+J_i^{(\tau)}S_{i,\tau} S_{i,\tau +1})\nonumber\\
&&-\sum_{i,\tau ,\tau\sp{\prime}} K_{\tau ,\tau\sp{\prime}} S_{i,\tau} S_{i,\tau\sp{\prime}}.
\label{H_classical}
\end{eqnarray}
Here, $S_{i,\tau}$ is a classical XY spin (i.e., a two-component unit vector) at position $i$ in space and $\tau$ in imaginary time. $J_i^{(s)}$ and $J_i^{(\tau)}$ are (random) ferromagnetic interactions between nearest neighbor spins in space and imaginary time directions, respectively.
As the quenched disorder is time-independent, their values depend on the space coordinate $i$ but not on the imaginary time coordinate $\tau$ (i.e., the disorder is columnar or perfectly correlated in the time direction).
The long-range interaction $K_{\tau,\tau\sp{\prime}}$ in the time direction arises from the dissipative dynamics of the quantum action.
It is given by
\begin{equation}
K_{\tau,\tau\sp{\prime}}=\gamma |\tau - \tau\sp{\prime}|^{-\alpha},\label{Long_interaction}
\end{equation}
where $\gamma$ is the  interaction amplitude while the exponent $\alpha$ takes the value 2 for Ohmic dissipation ($z_0=2$). The values
of $J_i^{(s)}, J_i^{(\tau)}$, and $K_{\tau,\tau\sp{\prime}}$ are determined by the parameters of the quantum action
(\ref{qaction}). However, as we are interested in the universal aspects of the phase transition only,
their precise values do not matter. We therefore tune the transition by varying the temperature $T$ of the classical XY model (\ref{H_classical}), while keeping $J_i^{(s)}, J_i^{(\tau)}$, and $K_{\tau,\tau\sp{\prime}}$ constant. This classical temperature $T$ differs
from the actual (physical) temperature $T_Q$ of the quantum system which maps onto the inverse system size in imaginary time direction, $L_\tau^{-1}$, of the XY model (\ref{H_classical}).
Under the quantum-to-classical mapping from the action (\ref{qaction}) to the Hamiltonian (\ref{H_classical}), the superfluid density of the quantum system maps onto the spin-wave stiffness in space direction, and the compressibility maps onto the stiffness in imaginary time direction.

\section{Theory}
\label{sec:Theory}
\subsection{Renormalization group predictions}
\label{subsec:Theory_RR}

Hoyos et al.\ \cite{HoyosKotabageVojta07,VojtaKotabageHoyos09} performed a strong-disorder renormalization group analysis of the LGW theory (\ref{qaction}) with quenched disorder in the large-$N$ limit. This analysis yielded a quantum critical point of exotic infinite-randomness type that belongs to the random transverse-field Ising chain universality class \cite{Fisher92,Fisher95}. Whereas the dynamical scaling in the absence of disorder is of power-law type, the disordered system features unconventional activated dynamical scaling characterized by an exponential
relation between correlation time and length, $\text{ln} \xi_\tau \sim \xi^\psi$ with $\psi=0.5$.

Specifically, the strong-disorder renormalization group  makes the following predictions for the finite-size scaling behavior of observables (see also Refs. \cite{MohanNarayananVojta10,HrahshehBarghathiVojta11}).
Right at criticality, the order parameter susceptibility  $\chi $ is expected to depend on the system size $L_\tau$ in imaginary time direction via
\begin{equation}
\chi \sim L_\tau[\text{ln}(L_\tau/b)]^{2\phi-1/\psi},\label{Griff_0}
\end{equation}
where $\phi=(1+\sqrt{5})/2$ and $\psi=1/2$ are the cluster size and tunneling critical exponents of the infinite-randomness critical point, respectively ($b$ is an arbitrary microscopic scale). The logarithmic $L_\tau$ dependence in (\ref{Griff_0}) reflects the activated dynamical scaling.
In the ordered Griffiths phase ($T<T_c$), the susceptibility diverges as
\begin{equation}
\chi \sim L_\tau^{1+1/z},\label{Griff_1}
\end{equation}
and in the disordered Griffiths phase ($T>T_c$), it behaves as
\begin{equation}
\chi \sim L_\tau^{1-1/z}.\label{Griff_2}
\end{equation}
The nonuniversal Griffiths dynamical exponent $z$ varies with temperature.
Upon approaching the critical point, it diverges as
\begin{equation}
z \sim |T-T_c|^{-\nu\psi}.\label{Griff_3}
\end{equation}
where $\nu=2$ is the correlation length exponent of the infinite-randomness critical point.

The spin-wave stiffness $\rho_s$ of the XY Hamiltonian describes the change in the free-energy density $f$ during a twist of the spins at two opposite boundaries by an angle $\theta$. For small $\theta$ and large system size $l$, the free-energy change reads
\begin{equation}
f(\theta)-f(0)=\frac{\rho_s}{2}\frac{\theta^2}{l^2}. \label{free_energy_stiff}
\end{equation}
We need to distinguish  two kinds of stiffnesses, the space-stiffness $\rho_s^{(s)}$, and the time-stiffness $\rho_s^{(\tau)}$. To discuss the space-stiffness $\rho_s^{(s)}$ (which corresponds to the superfluid density of the original quantum system), we implement the twist between $0$ and $L$ in space direction with $l=L$ in Eq.~(\ref{free_energy_stiff}). For the time-stiffness $\rho_s^{(\tau)}$ (which corresponds to the compressibility of the quantum system), the twisted boundary conditions are between 0 and $L_\tau$ in imaginary-time direction, and $l=L_\tau$.

\begin{figure}
\centering
\includegraphics[width=7.5cm ]{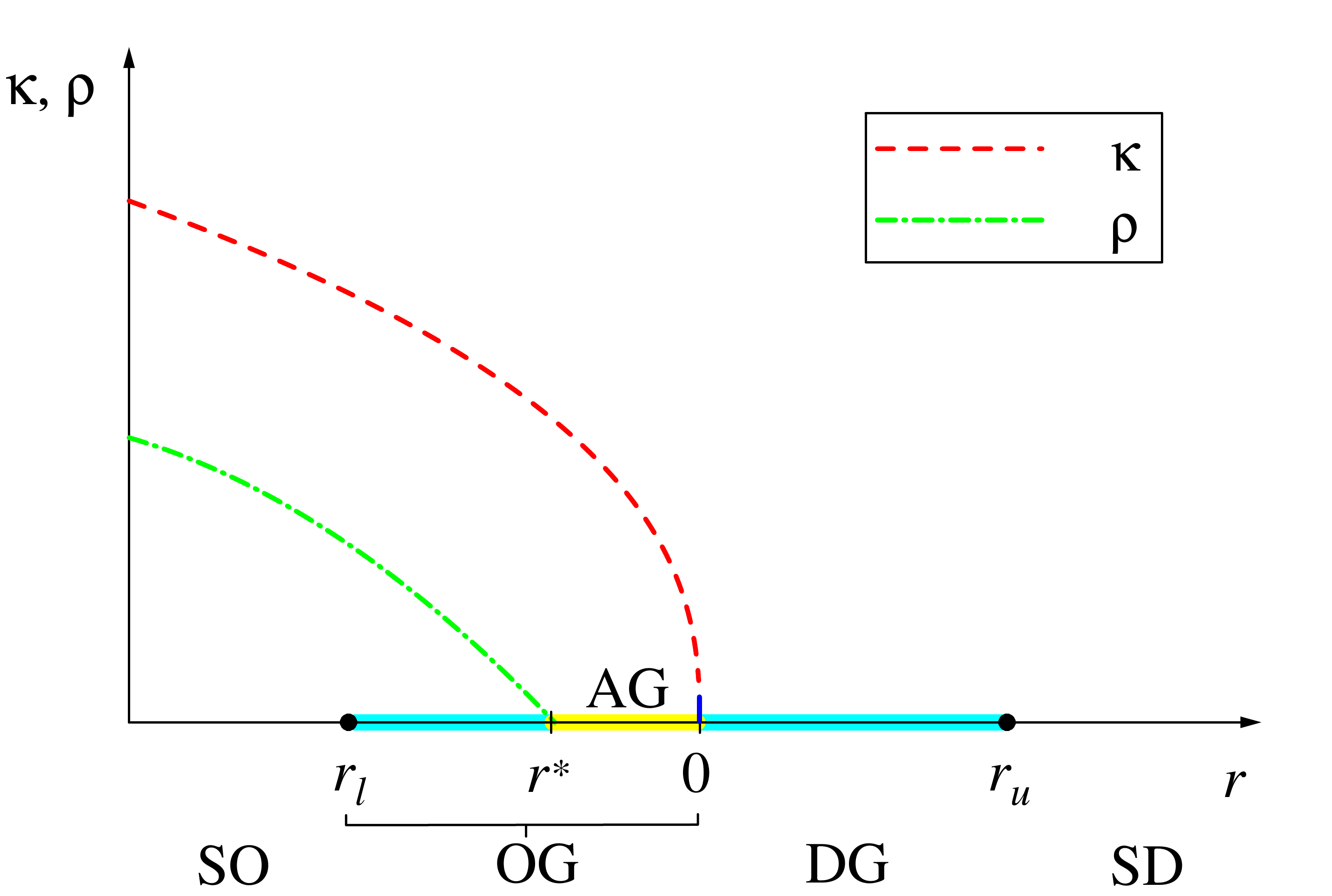}
\caption{Schematic behavior of the compressibility $\kappa$ and the superfluid density $\rho$ near the quantum phase transition.
$r$ denotes the distance from criticality. In the classical XY model (\ref{H_classical}), $\kappa$ and $\rho$ are represented by the spin-wave stiffnesses  $\rho_s^{(\tau)}$ and $\rho_s^{(s)}$, respectively, and $r\sim T-T_c$. The strongly ordered and disordered conventional phases are marked by SO and SD whereas OG and DG are the ordered and disordered Griffiths phases. The superfluid density vanishes in the anomalous part (AG) of the ordered Griffiths phase.}
\label{fig:stiff_schematic}
\end{figure}

Both stiffnesses vanish in the disordered phase above $T_c$. In a conventional phase transition scenario, they would be expected to be nonzero everywhere in the long-range ordered phase below $T_c$. However,
the theory developed in Refs.~\cite{MohanNarayananVojta10,MGNTV10} implies anomalous behavior of $\rho_s^{(s)}$: Because the distribution of the effective interactions $J_{eff}^{(s)}$ becomes very broad under the renormalization group, the stiffness vanishes for $L\to \infty$ in part of the ordered Griffiths phase between $T_c$ and $T^*$ where $T^*$ is the temperature where the Griffiths dynamical exponent $z=1$ (see schematic Fig.\
\ref{fig:stiff_schematic}).
In this temperature range, it behaves as
\begin{equation}
\rho_s^{(s)} \sim L^{1-z}.
\label{stiffness_space}
\end{equation}
Normal behavior with nonzero $\rho_s^{(s)}$ in the thermodynamic limit is restored for $T<T^*$.
In contrast, the time-stiffness $\rho_s^{(\tau)}$ is nonzero everywhere in the ordered phase, and behaves as
\begin{equation}
\rho_s^{(\tau)} \sim |T-T_c|^\beta,
\label{stiffness_time}
\end{equation}
where $\beta=2-\phi$ is the exponent of the order parameter.

\section{Monte Carlo simulations}
\label{sec:MC}

To confirm the predictions of the strong-disorder renormalization group summarized in Sec. \ref{sec:Theory}, we perform extensive Monte Carlo simulations of the (1+1)-dimensional XY Hamiltonian (\ref{H_classical}) with long-range interactions in time direction.

We employ the Luijten algorithm \cite{LuijtenBlote95}, a version of the Wolff cluster algorithm \cite{Wolff89} that is optimized for long-range
interactions. Its numerical effort (per sweep) scales linearly with the total number of sites $N_s=L\, L_\tau$ (rather than $L \, L_\tau^2$ as expected
in a naive implementation of the Wolff algorithm for long-range interactions).
Using this method, we simulate systems with linear size $L=20$ to 160 in the space direction and $L_\tau =2$ to 40000 in the (imaginary) time direction.
To introduce quenched disorder that is perfectly correlated in the imaginary time direction, we treat the interactions $J_i^{(s)}$ and $J_i^{(\tau)}$ as independent random variables. Details of their
probability distributions are unimportant for the universal properties we are interested in. We thus employ a simple binary probability distribution
\begin{equation}
\rho(J)=(1-c)\delta(J-J_l)+c\delta(J-J_h) \label{Binary}
\end{equation}
having a higher value $J_h$ with concentration $c$ and a lower value $J_l$ with concentration $(1-c)$. This binary distribution is numerically efficient and 
allows us to independently control the strength and concentration of the defects. In most of the simulations, we choose parameters $c=0.5$,  $J_h$=2 and $J_l$=0.5, and an interaction amplitude $\gamma =0.1$
of the long-range temporal interaction.

For each simulation run we perform 100 Monte Carlo sweeps for equilibration (one sweep is defined as a number of cluster flips such that the total number of flipped spins equals the total number $N_s=L\,L_t$ of sites). We have confirmed that this is sufficient by comparing the results of runs with hot starts (spins initially pointing in random directions) and cold starts (spins inially aligned). We then perform 200 sweeps during which we measure the energy, specific heat, order parameter, susceptibility, Binder cumulant, correlation function, correlation length as well as the space and time stiffnesses (with a measurement taken after every sweep). All observables are averaged over 1000 to 10 000 disorder configurations. Using short measurement runs for a large number of disorder configurations improves the overall performance, as is discussed in Refs.\ \cite{BFMM98,BFMM98b,VojtaSknepnek06,ZWNHV15}.

\section{Results}
\label{view}

\subsection{Clean system}
To test our simulation method and to make contact with the literature, we first analyze the clean case with uniform $J_i^{(s)}=J_i^{(\tau)}=1$.
We compute the order parameter
\begin{equation}
m=\frac{1}{N_s}\sum_{i,\tau} S_{i,\tau}.
\end{equation}
and its Binder cumulant \cite{Binder81b,BinderLandau84}
\begin{equation}
g=1-\frac{\langle m^4\rangle}{3\langle m^2\rangle^2}~.
\end{equation}
We use the finite-size scaling method \cite{Barber_review83,Cardy_book88} to estimate the location of the critical point.
The Binder cumulant is expected to have a finite-size scaling form of
\begin{equation}
g(r,L,L_\tau)=\Upsilon_g(rL^{1/\nu_{cl}},L_\tau/L^{z_{cl}}).\label{scaling_Binder}
\end{equation}
Here, $r=T-T_c$ is the distance from the critical point, $\Upsilon_g$ is the scaling function, $\nu_{cl}$ is the clean correlation length exponent, and $z_{cl}$ is the clean dynamical exponent.

The long-range interactions break the symmetry between the space and imaginary time directions, implying that the spatial and temporal system sizes
$L$ and $L_\tau$ scale differently.  The value of $z_{cl}$ is therefore not known at the outset, and we need to perform anisotropic finite-size scaling. We employ the iterative method outlined in
Refs.~\cite{GuoBhattHuse94,RiegerYoung94,SknepnekVojtaVojta04,Vojtaetal16}: The Binder cumulant has a maximum as function of $L_\tau$ for fixed $L$. According to (\ref{scaling_Binder}), the peak position $L_\tau^{max}$ must behave as $L^{z_{cl}}$ at criticality, and the value $g^{max}$ of the maximum must be $L$-independent. The sizes $L$ and $L_\tau^{max}$ in the space and imaginary-time directions define the ``optimal shape''
for finite-size scaling.

\begin{figure}
\centering
\includegraphics[width=8.2cm ]{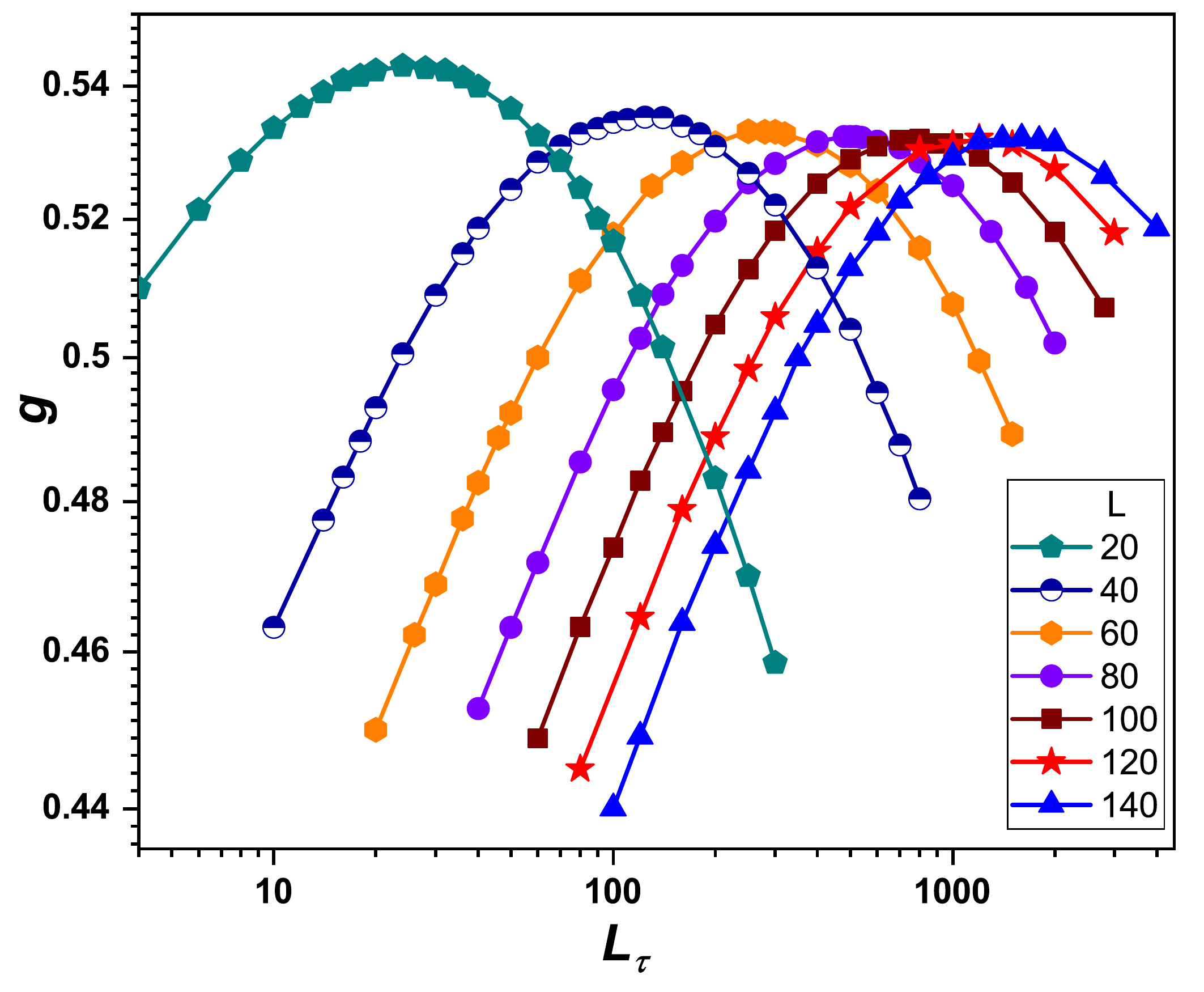}
\caption{Binder cumulant $g$ of the clean system vs.\ $L_\tau$ for different $L$ at the critical temperature $T_c=0.56969$. The interactions
are uniform, $J_i^{(s)}=J_i^{(\tau)}=1$ and $\gamma=0.1$. The statistical error of $g$ is much smaller than the symbol size.}
\label{fig_Binder_curves}
\end{figure}

\begin{figure}
\centering
\includegraphics[width=8.2cm ]{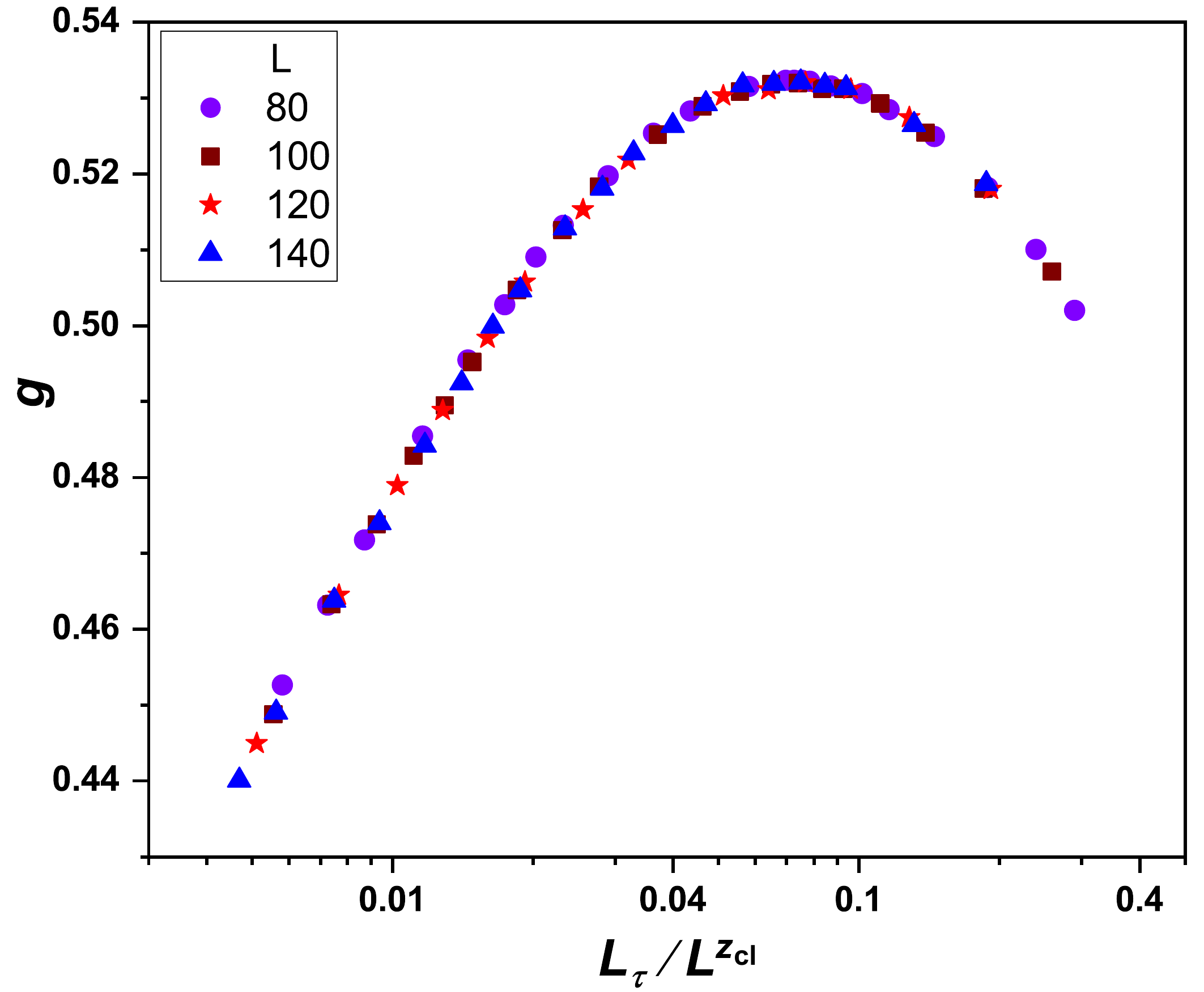}
\caption{Scaling plot of the Binder cumulant data from Fig.\ \ref{fig_Binder_curves}. It shows
$g$ as a function of  $L_\tau/L^{z_{cl}}$ for different $L$ at $T_c=0.56969$. All curves collapse and follow the power-law scaling
form (\ref{scaling_Binder}) for $z_{cl}=2.01$.}
\label{fig_Binder_scale}
\end{figure}

Fig.~\ref{fig_Binder_curves} shows the behavior of the Binder cumulant $g$ as a function of $L_\tau$ for several system sizes $L$ at the estimated critical temperature $T_c=0.56969(6)$. These curves indeed have identical maximum values that are independent of the system size $L$ (because of corrections to scaling, some deviations occur at small $L$).
To test the scaling form of the Binder cumulant (\ref{scaling_Binder}) and to measure the clean dynamical exponent $z_{cl}$, we plot the data of the Binder cumulant as a function of $L_\tau/L^{z_{cl}}$ and vary $z_{cl}$ until a good collapse is achieved. As shown in Fig.~\ref{fig_Binder_scale}, the data collapse onto each other for a dynamical exponent value $z_{cl}=2.01(6)$.
This also implies that the dynamical scaling is of conventional power-law form, as expected \cite{PFGKS04,WernerTroyerSachdev05}.

\begin{figure}
\centering
\includegraphics[width=8.2cm ]{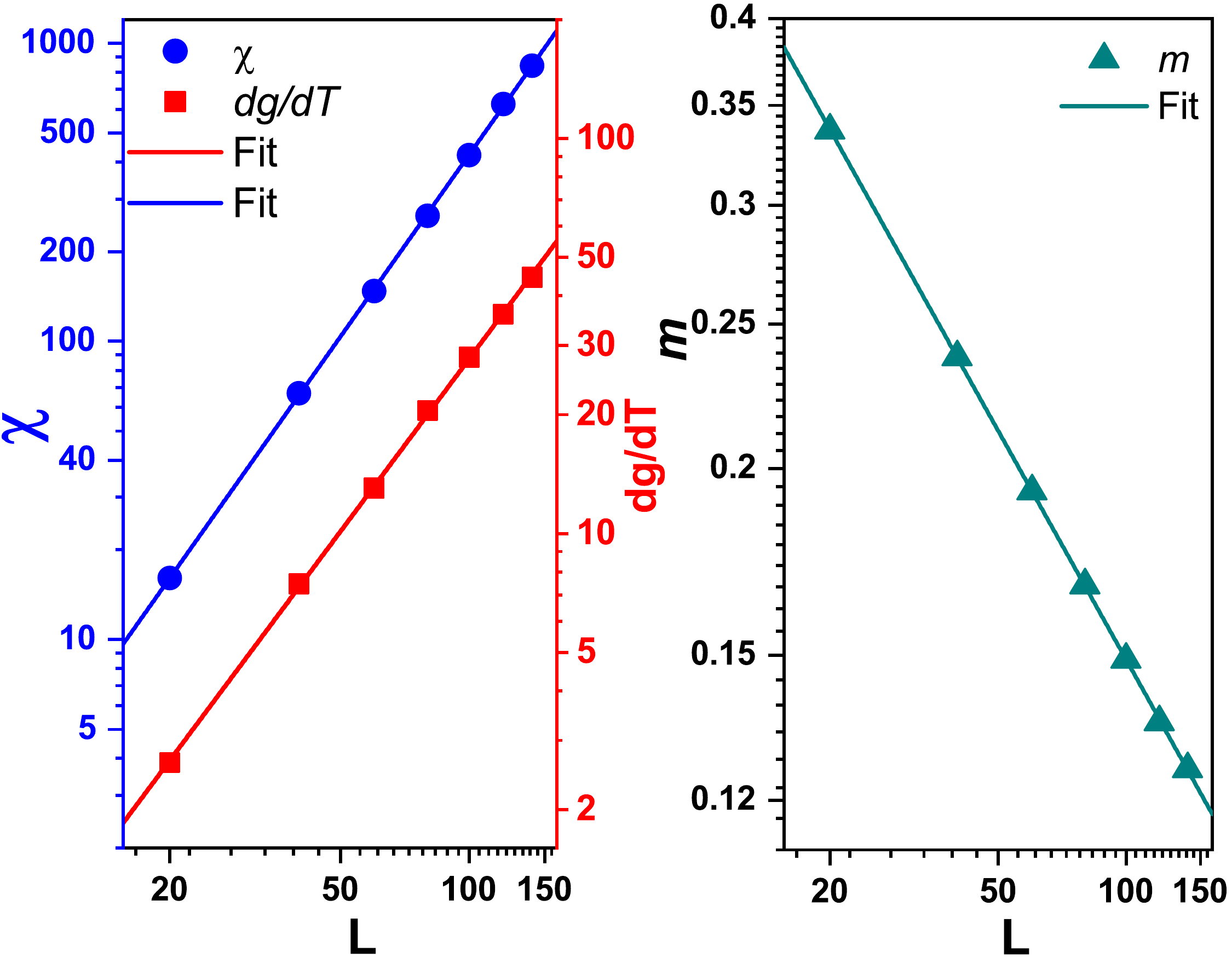}
\caption{Left: Double logarithmic plot of susceptibility $\chi$ and slope of Binder cumulant $dg/dT$ vs. system size $L$ for optimally shaped clean samples at criticality. The solid lines are fits to the predicted power laws $\chi \sim L^{\gamma/\nu}$ and $dg/dT\sim L^{1/\nu_{cl}}$ with $\gamma_{cl}/\nu_{cl}=2.03$ and $1/\nu_{cl}=1.454$. Right: Magnetization $m$ as a function of $L$ at criticality. The slope of the fitted line with $m\sim L^{-\beta_{cl}/\nu}$ gives $\beta_{cl}/\nu_{cl}=0.507$.}
\label{fig_Binder_slope}
\end{figure}

To compute further critical exponents of the system, we analyze the properties of the order parameter, its susceptibility, and the slope $dg/dT$ of the Binder cumulant at the critical temperature. The correlation length exponent $\nu$ can be estimated from the slopes of Binder cumulant. Taking the derivative with respect to temperature in (\ref{scaling_Binder}), it follows that $dg/dT$ at criticality $(r=0)$ behaves as $L^{1/\nu_{cl}}$ (if evaluated for the optimal sample shapes $L_\tau=L_\tau^{max} \sim L^{z_{cl}}$). In Fig.~\ref{fig_Binder_slope}, we plot the slopes as a function of system size $L$. The critical exponent $\nu_{cl}=0.687(9)$ follows from a power-law fit of these data to $dg/dT\sim L^{1/\nu_{cl}}$.
We also study the system-size dependence of the order parameter and its susceptibility at the critical temperature, as shown in
Fig.~\ref{fig_Binder_slope}.
We find that these observables feature the expected power-law behavior (if evaluated for the optimal sample shapes $L_\tau=L_\tau^{max} \sim L^{z_{cl}}$),  $m\sim L^{-\beta_{cl}/\nu_{cl}}$ and $\chi \sim L^{\gamma_{cl}/\nu_{cl}}$ \cite{Barber_review83}. Using power-law fits, we estimate the critical exponents to be $\beta_{cl}/\nu_{cl}=0.507(5)$ and $\gamma_{cl}/\nu_{cl}=2.03(4)$.

The exponents $\beta_{cl}/\nu_{cl}$, $\gamma_{cl}/\nu_{cl}$, and $z_{cl}$ are not independent of each other. They are connected by the hyper-scaling relation \cite{ContinentinoJapiassuTroper89}
\begin{equation}
\frac{2\beta_{cl}}{\nu_{cl}}+\frac{\gamma_{cl}}{\nu_{cl}}=d+z_{cl}
\end{equation}
where $d=1$ is the number of space dimensions. We find that our numerical estimates fulfill the hyper-scaling relation within their error bars.
Moreover, all clean exponents agree with those found in Ref.~\cite{WernerTroyerSachdev05}.

Note that the correlation length exponent $\nu_{cl}=0.687$ violates the Harris criterion $d \nu_{cl} > 2$ \cite{Harris74}, implying that the clean critical behavior will be unstable against disorder.

\subsection{Disordered system}

After studying and analyzing the behavior of the clean critical point, we now apply quenched disorder to the Hamiltonian (\ref{H_classical}) by making the ferromagnetic interactions  $J_i^{(s)}$ and $J_i^{(\tau)}$ random functions of the space coordinate $i$, drawn from the
probability distribution (\ref{Binary}). We have attempted to use the same Binder-cumulant-based finite-size scaling method as in the clean case to find the critical point. Unfortunately, this analysis is hampered by strong corrections to scaling. 
We instead use finite-size scaling in just $L_\tau$ to study the order parameter susceptibility $\chi$ and the correlation time $\xi_\tau$, in analogy to Ref.~\cite{HrahshehBarghathiVojta11} where similar difficulties were encountered. This requires samples having effectively infinite size $L$ in the space direction ($L \gg L_\tau$). To analyze $\chi$ and $\xi_\tau$, we have therefore simulated systems of fixed spatial size $L=1000$ and varied $L_\tau$ from 10 up to 448. In contrast, the spatial correlation function, the superfluid density, and the compressibility are measured in systems of fixed size closed to the optimal shapes (where the correlations
extend equally in the space and time directions). 
 
\begin{figure}
\centering
\includegraphics[width=8.2cm ]{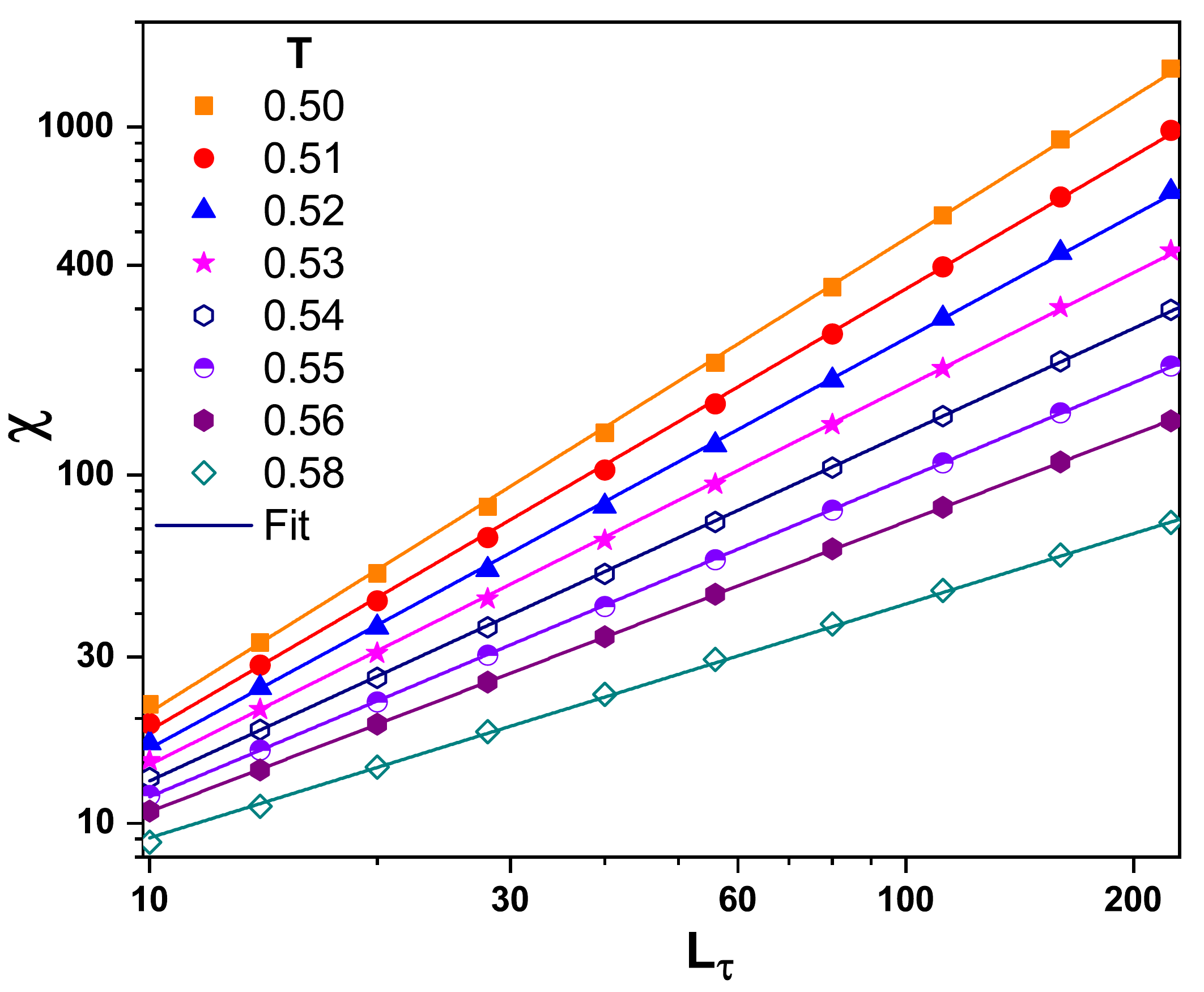}
\caption{Susceptibility $\chi$ of the disordered system as a function of $L_\tau$ for various temperatures in the Griffiths phase. The size in space direction is $L=1000$. The solid lines are fits to the power laws (\ref{Griff_1}) and (\ref{Griff_2}).}
\label{fig_SUS_scale}
\end{figure}

\begin{figure}
\centering
\includegraphics[width=8.2cm ]{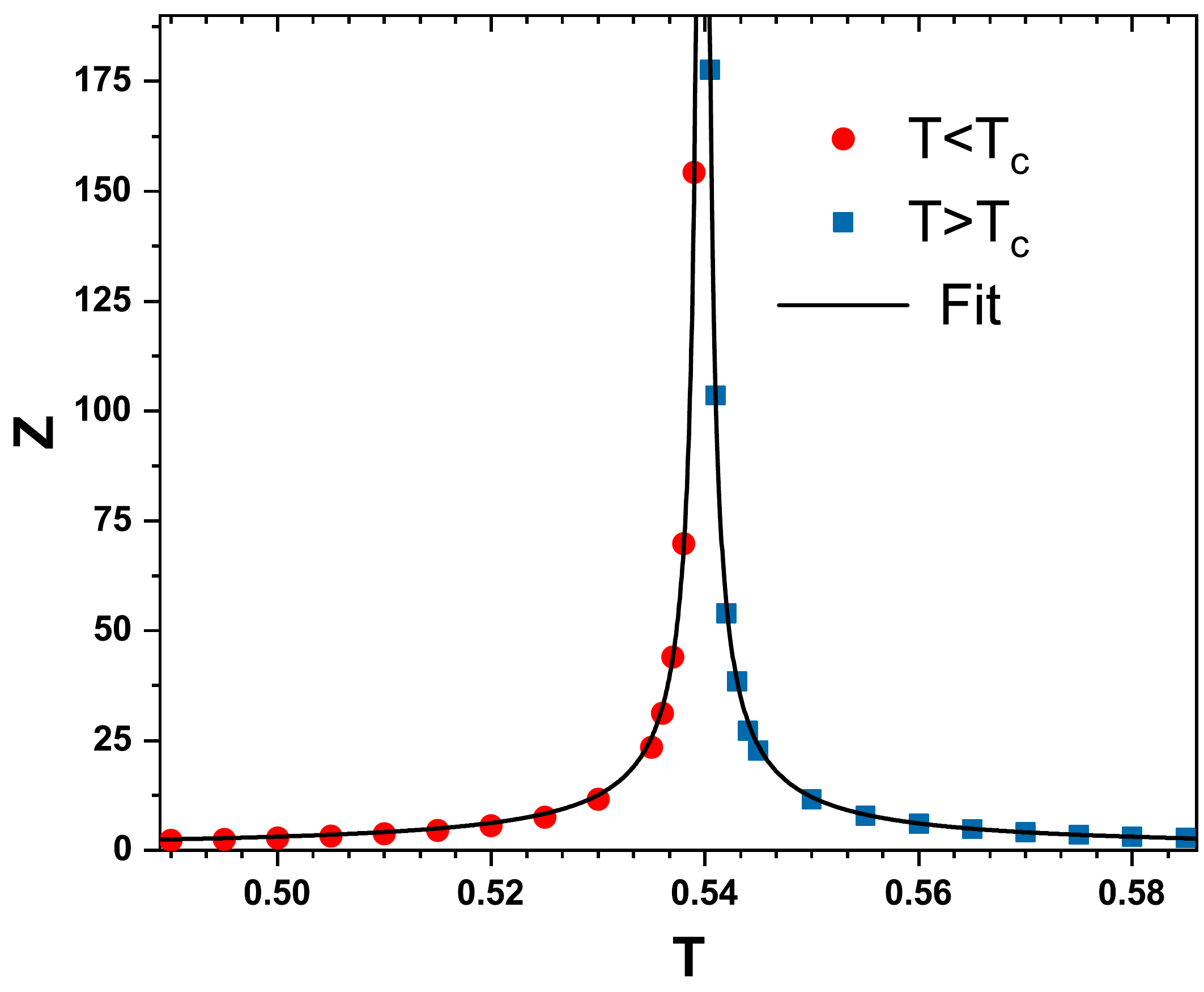}
\caption{Dynamical exponent $z$ as a function of classical temperature $T$ in Griffiths regions. The data are extracted from the susceptibility data in Fig.~\ref{fig_SUS_scale}. The solid lines are fits of $z$ to Eq.~(\ref{z_exponent}).}
\label{fig_z_scale}
\end{figure}

Fig.~\ref{fig_SUS_scale} presents the order parameter susceptibility $\chi$, confirming that $\chi$ follows power laws in $L_\tau$ for an entire range of temperatures (the Griffiths region), as predicted theoretically in eqs.\  (\ref{Griff_1}) and (\ref{Griff_2}).
The Griffiths dynamical exponent $z$ can be determined by fitting the susceptibility to eqs.~(\ref{Griff_1}) and (\ref{Griff_2}). Its value is predicted to vary with temperature and to diverge at the critical point as
\begin{equation}
z \sim \frac{1}{|T-T_c|}. \label{z_exponent}
\end{equation}
Fig.~\ref{fig_z_scale} shows the values of $z$ in the ordered and disordered Griffiths phase as a function of temperature $T$. These values are fitted to the power law relation (\ref{z_exponent}) giving an estimated critical temperature of $T_c\approx 0.540(9)$.

In addition to the order parameter susceptibility, we also investigate the superfluid density and the compressibility of the quantum system. In the classical problem (\ref{H_classical}), they are represented by the spin-wave stiffnesses in the space and time directions, respectively.
The spatial stiffness can be computed using the relation \cite{TeitelJayaprakash83}
\begin{eqnarray}
\rho_s^{(s)}&=&\frac{1}{N_s}\sum_{i,j,\tau,\tau\sp{\prime}} J_{i,j,\tau,\tau\sp{\prime}} \langle \vec{S}_{i,\tau} \cdot \vec{S}_{j,\tau\sp{\prime}} \rangle (i-j)^2\nonumber\\
&&-\frac{1}{N_s T}\Big \langle \Big( \sum_{i,j,\tau,\tau\sp{\prime}} J_{i,j,\tau,\tau\sp{\prime}} \mathbf{\hat{K}} \cdot( \vec{S}_{i,\tau} \times \vec{S}_{j,\tau\sp{\prime}} ) (i-j)\Big)^2 \Big\rangle,\nonumber\\\label{stiff_space}
\end{eqnarray}
where
\[
   J_{i,j,\tau,\tau\sp{\prime}}=
  \left\{ {\begin{array}{cll}
   J_i^{(s)}&\text{if}~~j=i\pm 1~,&\tau=\tau\sp{\prime} \\
   J_i^{(\tau)} & \text{if}~~ i=j~~~~~~, & \tau=\tau\sp{\prime}\pm 1~~~, \\
   \gamma{|\tau-\tau\sp{\prime}|^{-\alpha}} & \text{if}~~ i=j~~~~~~, & \tau\neq\tau\sp{\prime}~.\\
   0 & \text{otherwise} \\
  \end{array} } \right.
\]
Here, $N_s=LL_\tau$ is the total number of sites, and $\mathbf{\hat{K}}$ is the unit vector perpendicular to the $xy$-plane in spin space.
For the calculation of $\rho_s^{(\tau)}$, the term $(i-j)$ has to replaced by $(\tau-\tau\sp{\prime})$.

\begin{figure}
\centering
\includegraphics[width=8.2cm ]{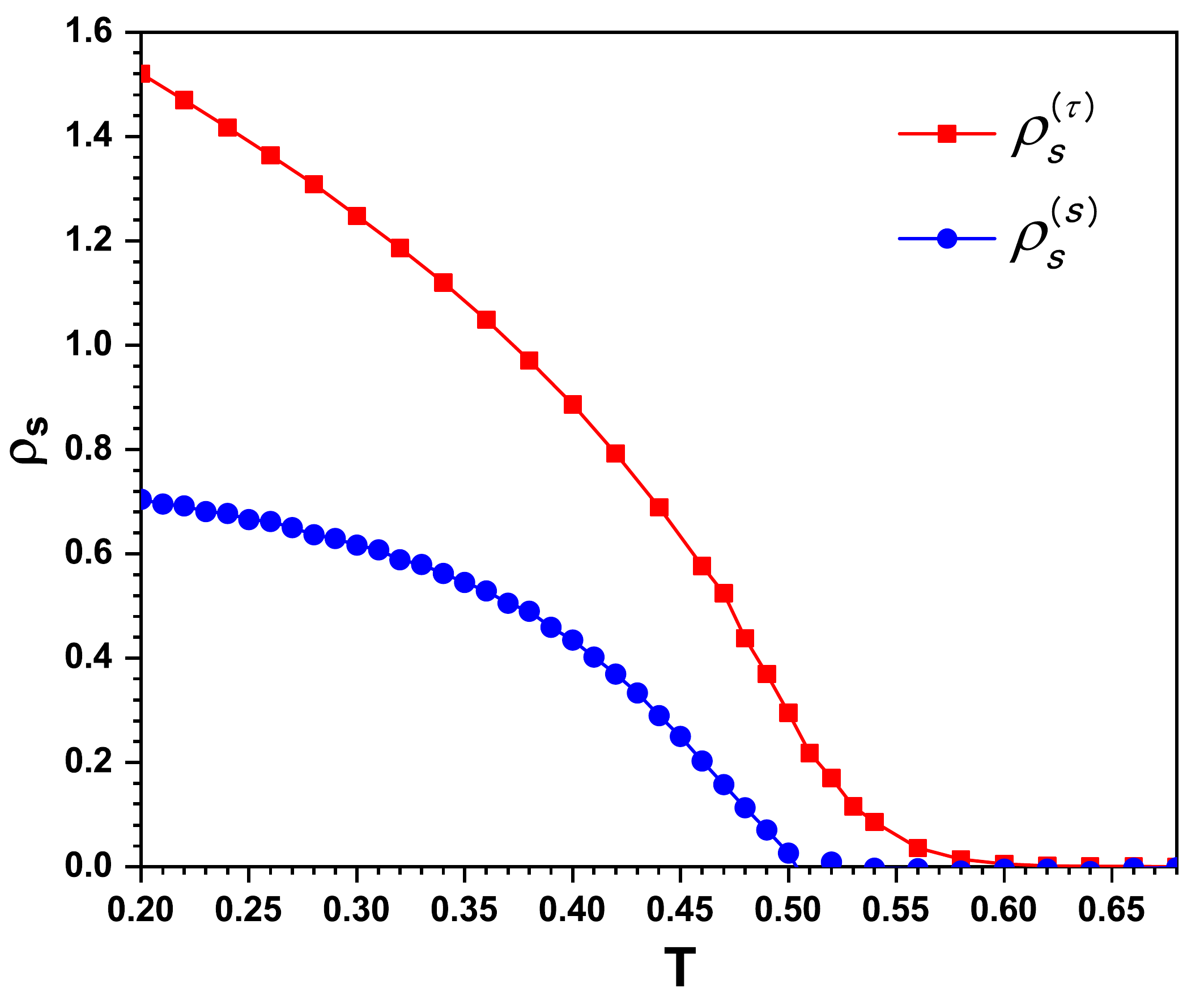}
\caption{Spin-wave stiffness in space $\rho_s^{(s)}$ and time $\rho_s^{(\tau)}$ as function of the classical temperature $T$ with system  size $L=160$ and $L_\tau=10000$. The data for $\rho_s^{(s)}$ are rescaled by 500 for clarity.}
\label{fig_stiffness}
\end{figure}

The behavior of the spin-wave stiffnesses is illustrated in Fig.~\ref{fig_stiffness}. It shows the results for the space and time stiffness of a system of size $L=160$ and $L_\tau=10000$. Clearly, the two stiffnesses behave differently. According to Eq.~(\ref{stiffness_time}), the imaginary-time stiffness (i.e., the compressibility of the original quantum system) is expected to behave as $|T-T_c|^\beta$, i.e., it vanishes
at $T_c$. Despite significant finite-size rounding, our data are qualitative compatible with this prediction, giving $T_c\approx 0.53$, in agreement with our earlier estimate of $T_c = 0.540(9)$. In contrast, the space stiffness $\rho_s^{(s)}$ is more than two orders of magnitude smaller even though the microscopic interaction strengths do not have a significant anisotropy. Moreover, $\rho_s^{(s)}$ vanishes at a lower temperature $T^* \approx 0.50$, giving rise to anomalous elasticity \cite{MGNTV10} for temperatures between $T_c$ and $T^*$.

In addition, we compute the correlation function $G$ in the space and time directions to determine the corresponding correlation lengths
$\xi$ and $\xi_\tau$, respectively. This allows us to estimate the correlation length exponent $\nu$. The correlation functions in space and time directions are defined as
\begin{eqnarray}
G(x) &=& \frac{1}{N_s}\sum_{i,j,\tau} \langle \vec{S}_{i,\tau} \cdot \vec{S}_{j,\tau} \rangle \delta (x-|i-j|),\label{Cspace_function}\\
G(\tau)&=&\frac{1}{N_s}\sum_{i,\tau_1,\tau_2} \langle \vec{S}_{i,\tau_1} \cdot \vec{S}_{i,\tau_2} \rangle \delta (\tau-|\tau_1-\tau_2|).
\end{eqnarray}

\begin{figure}
\centering
\includegraphics[width=8.2cm ]{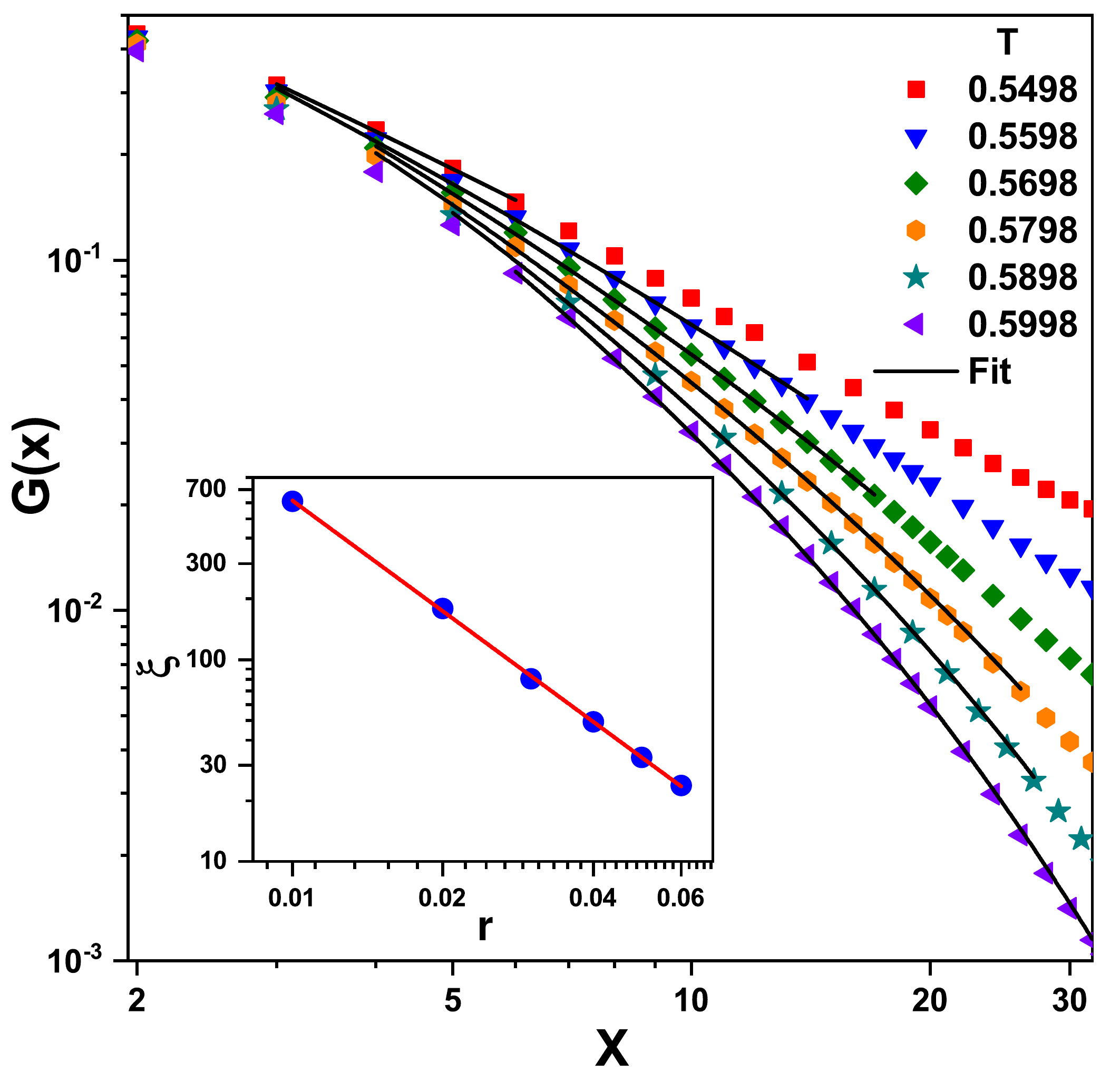}
\caption{Space-correlation function $G(x)$ for several temperature in Griffiths phase. The solid lines are fits to Eq.~(\ref{corr_function_space}). Inset: The space-correlation length $\xi$ obtained by analyzing space-correlation function as a function of distance $\delta$ from critical temperature. The solid line is a fit of Eq.~(\ref{c_length_pwl}).}
\label{fig_correlation_function}
\end{figure}

Fig.~\ref{fig_correlation_function} shows the spatial correlation function (\ref{Cspace_function}) for different temperatures in the Griffiths region above $T_c$ for a system of sizes $L=80$ and $L_\tau=1200$.
Because of the small spatial system size, there are significant finite-size effects for temperatures close to $T_c=0.54$
for which the correlations decay slowly (notice the unexpected upturns for large $|\mathbf{x}|$ of the correlation functions
for the lower temperatures).
The strong-disorder renormalization group \cite{Fisher95} predicts that the spatial correlation function behaves as
\begin{equation}
G(x)\sim \frac{\text{exp}[-(x/\xi)-(27\pi^2/4)^{1/3}(x/\xi)^{1/3}]}{(x/\xi)^{5/6}} ~\label{corr_function_space}~.
\end{equation}
The figure shows that $G(x)$ is well described by this functional form for temperatures away from $T_c$.
Closer to $T_c$, the agreement becomes questionable, because the finite-size effects restrict the fits to a very narrow
$|\mathbf{x}|$-range. Ignoring these complications, we can extract values of $\xi$ by fitting the $G(x)$ data to Eq.~(\ref{corr_function_space}) for distances between $x=3$ and some cutoff at which the curves start to become noisy or deviate from the expected behavior.
The inset of Fig.~\ref{fig_correlation_function} shows the relation between the correlation length $\xi$ and the distance from criticality $r=T-T_c$ which reads \cite{Fisher95}
\begin{equation}
\xi \sim |r|^{-\nu}.\label{c_length_pwl}
\end{equation}
As expected, the data can be fitted to the power law (\ref{c_length_pwl}), giving a correlation length exponent of $\nu=1.8(3)$, in reasonable agreement with the renormalization group result $\nu=2$ \cite{Fisher95}.

We also analyze the average correlation time $\xi_\tau$. As we can reach much larger temporal system sizes $L_\tau$
than spatial system sizes
 $L$, we can compute $\xi_\tau$ via the standard second-moment method \cite{CooperFreedmanPreston82,Kim93,CGGP01} from the Fourier transform $\tilde{G}(\omega)$ of the temporal correlation function $G(\tau)$:
\begin{equation}
\xi_\tau=\left[\frac{\tilde{G}(0)-\tilde{G}(\omega_{min})}{\omega_{min}^2\tilde{G}(\omega_{min})}\right]^{1/2}.
\end{equation}
Here, $\omega_{min}$ is the minimum wave number, $2\pi/L_\tau$ in the imaginary-time direction.

\begin{figure}
\centering
\includegraphics[width=8.2cm ]{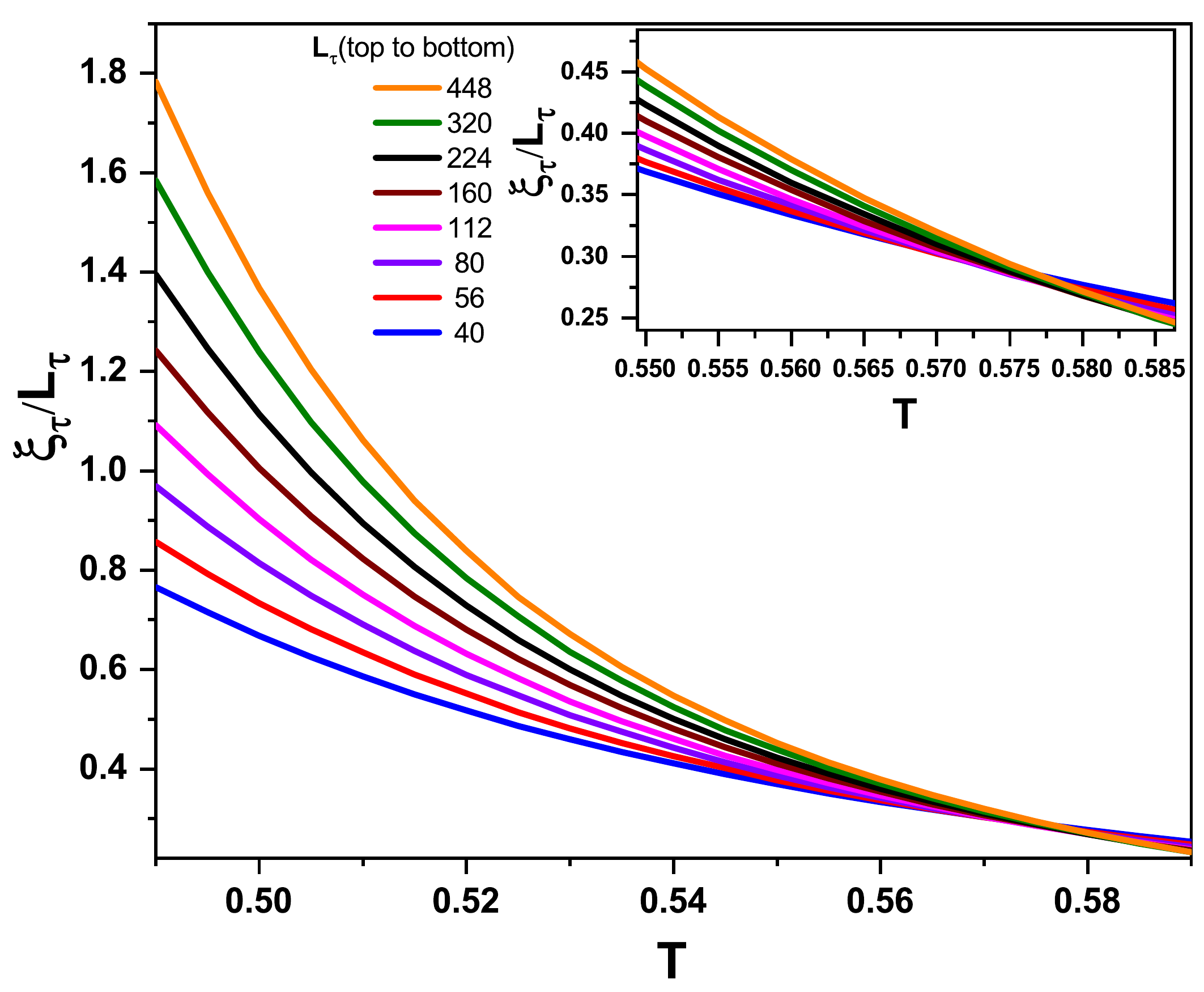}
\caption{Scaled correlation time $\xi_\tau/L_\tau$ versus temperature $T$ for different values of $L_\tau$ in the Griffiths region. The system size in space is $L=1000$; the data are averaged over 2000 disorder configurations. The inset shows a magnification for the crossing point of the curves.}
\label{fig_correlation_time_scale}
\end{figure}

The behavior of the correlation time $\xi_\tau$ in the Griffiths phase is illustrated by plotting $\xi_\tau/L_\tau$ as a function of temperature $T$ for several system of size $L_\tau$, as shown in Fig.~\ref{fig_correlation_time_scale}.
Remarkably, the different curves cross at temperature $T\approx 0.578$, clearly much higher than our critical temperature $T_c\approx 0.54$. This indicates that the correlation time $\xi_\tau$ diverges in part of the disordered phase before reaching the phase transition. This behavior
is indicative of infinite-randomness physics and can be understood by estimating the rare region contribution to the correlation time $\xi_\tau$ \cite{HrahshehBarghathiVojta11}. It reads
\begin{equation}
\xi_\tau \sim \int\limits_{0}^{\epsilon_0} d\epsilon~ \epsilon^{1/z-1}~ \frac{1}{\epsilon},\label{corr_time_rare}
\end{equation}
where $\epsilon$ is the renormalized distance from criticality of rare region and the factor $\epsilon^{1/z-1}$
stems from the rare-region density of states. According to Eq.~(\ref{corr_time_rare}), the integral diverges for $z>1$ and converges for $z<1$. Therefore, the correlation time is expected to diverge in the disordered Griffiths phase (before reaching the phase transition) at the temperature at which the Griffiths dynamical exponent is $z=1$.
\footnote{The agreement between the crossing temperature in Fig.~\ref{fig_correlation_time_scale}, and the temperature at
which the susceptibility-based $z$-value (see Fig.~\ref{fig_z_scale}) equals 1 is not particularly good. This can be attributed to strong corrections to scaling, manifest, e.g., in the drift with $L_\tau$ of the crossing in
Fig.~\ref{fig_correlation_time_scale}.}

\section{Conclusions}
\label{sec:Conclusions}
In summary, we have studied the superconductor to metal quantum phase transition in ultra thin nanowires by performing large-scale Monte Carlo simulations. To this end, we have mapped the quantum action onto a (1+1)-dimensional classical XY Hamiltonian with long-rang interactions and columnar disorder.

 For the clean system, we have employed finite-size scaling of the Binder cumulant to estimate the critical point and determine the universality class of the phase transition. Our results agree well with earlier Monte-Carlo simulations \cite{WernerTroyerSachdev05} as well as perturbative renormalization group results \cite{PFGKS04,SachdevWernerTroyer04}.
In particular, the dynamical scaling is of conventional power-law type $\xi_\tau \sim \xi^{z_{cl}}$.

In the presence of quenched disorder, our results provide strong evidence in support of the exotic infinite-randomness behavior that was predicted by the strong-disorder renormalization group approach \cite{HoyosKotabageVojta07} and the large-$N$ saddle point analysis \cite{DRMS08}. It features activated dynamical scaling, $\ln \xi_\tau \sim \xi^{\psi}$.
In particular, the critical behavior is compatible with the universality class of the random transverse-field Ising chain. This may appear surprising at first glance because the random transverse-field Ising model has discrete symmetry and no dissipation while our current problem has continuous XY symmetry and Ohmic dissipation. However, the behavior agrees with the general classification of disordered quantum phase transitions developed in Refs.~\cite{VojtaSchmalian05,Vojta06,Vojta13,VojtaHoyos14}.
In both systems, the rare regions are right at the lower critical dimension of the problem, putting the transitions into class $B$ \cite{VojtaHoyos14}. Moreover, both clean transitions violate the Harris criterion, implying that the disordered transitions are in subclass $B2$ which features infinite-randomness criticality \cite{VojtaHoyos14}.

Rare regions also lead to unusual properties in the Griffiths phases on both sides of the phase transition. Specifically, the superfluid density vanishes in part of the long-range ordered (superfluid) Griffiths phase, giving rise to an anomalous (sliding) Griffiths phase with
unusual elastic properties \cite{MGNTV10,PekkerRefaelDemler10}. On the disordered
side of the transition, rare regions lead to a divergence of the average correlation time before the transition is reached.

Our Monte Carlo simulations can in principle be generalized to compute further observables including the dynamical conductivity in the regime $\omega \gg T$. This would allow us to test the predictions of Ref.\ \cite{DRHV10}. Due to the large numerical effort, this remains a task for the future.

Recently, a pair-breaking superconductor-metal quantum phase transition has been observed in amorphous Mo-Ge nanowires \cite{Kimetal2018}. In these wires, the disorder turns out to be rather weak
and plays no role in the measured temperature range. Thus, the experimental results are well described by the clean theory of Refs.\ \cite{DRSS08,DelMaestroRosenowSachdev09}.
However, several superconducting quantum phase transitions in two-dimensional systems have been interpreted in terms of infinite-randomness critical behavior analogous to that
studied in the present paper. These include transitions of ultrathin Ga films~\cite{Xingetal15}, La\textsubscript{2}AlO\textsubscript{3}/SrTiO\textsubscript{3} interfaces~\cite{Shenetal18}, flakes of ZrNCl and MoS\textsubscript{2}~\cite{SaitoNojimaIwasa2018}, monolayer NbSe\textsubscript {2}~\cite{Xingetal17}, and InO films \cite{Lewellynetal18}.

This work was supported by the NSF under Grants No. DMR-1506152, PHY-1125915 and PHY-1607611. Thomas Vojta is thankful for the hospitality of the Kavli Institute for Theoretical Physics, Santa Barbara, and the Aspen Center for Physics where parts of the work were performed.

\section*{Author contribution statement}

T.V. conceived and coordinated the study. A.K.I. performed the Monte Carlo simulations, analyzed the
data, and created the figures. A.K.I and T.V. wrote the manuscript.

\bibliographystyle{spphys-tv}
\bibliography{../00Bibtex/rareregions}
\end{document}